\documentclass[amsmath,amssymb,amsbsy,reprint,prb,preprintnumbers,showpacs,superscriptaddress]{revtex4-2}
\usepackage{graphicx,color}
\usepackage{dcolumn}
\usepackage{bm}
\usepackage{braket}
\usepackage{mathtools}
\usepackage{ulem}
\usepackage[breaklinks,colorlinks=true,linkcolor=blue,urlcolor=blue,citecolor=blue]{hyperref}
\usepackage{times}
\usepackage{physics}
\usepackage{latexsym}
\usepackage{amsmath, amssymb}
\usepackage{mathtools}
\usepackage{multirow}

\newcommand{\be}{\begin{eqnarray}}
\newcommand{\ee}{\end{eqnarray}}

\renewcommand{\theequation}{\arabic{equation}}

\begin{document}

\title{Disordered purification phase transition in hybrid random circuits}
\date{\today}
\author{Kengo Anzai} 
\affiliation{Faculty of Science and Engineering, Akita University, Akita 010-8502, Japan}
\affiliation{Department of Applied Physics, Graduate School of Engineering, Tohoku University, Sendai 980-8579, Japan}
\author{Hiroaki Matsueda} 
\affiliation{Department of Applied Physics, Graduate School of Engineering, Tohoku University, Sendai 980-8579, Japan}
\affiliation{Center for Science and Innovation in Spintronics, Tohoku University, Sendai 980-8577, Japan}
\author{Yoshihito Kuno} 
\affiliation{Graduate School of Engineering Science, Akita University, Akita 010-8502, Japan}

\begin{abstract}
Noise is inevitable in realistic quantum circuits. It arises randomly in space. Inspired by spatial non-uniformity of the noise, we investigate the effects of spatial modulation on purification phase transitions in a hybrid random Clifford circuit. As an efficient observable for extracting quantum entanglement in mixed states, we employ many-body negativity. The behavior of the many-body negativity well characterizes the presence of the purification phase transitions and its criticality. We find the effect of spatial non-uniformity in measurement probability on purification phase transition. The criticality of the purification phase transition changes from that of uniform probability, which is elucidated from the argument of the Harris criterion. The critical correlation length exponent $\nu$ changes from $\nu < 2$ for uniform probability to $\nu > 2$ for spatially modulated probability. 
We further investigate a setting where two-site random Clifford gate becomes spatially (quasi-)modulated. We find that the modulation induces a phase transition, leading to a different pure phase where a short-range quantum entanglement remains.
\end{abstract}


\maketitle
\section{Introduction}
Measurement-induced phase transition (MIPT) is characterized by averaged entanglement entropy of pure states in random quantum circuits \cite{fisher_2019,Skinner2019,Fisher_annurev}. The circuits are composed of local measurements and random unitary gates, and the former (latter) elements break (induce) local entanglement of the initial state. The origin of the transition is such competition of entanglement. Besides, these gate operations can be viewed as an artificial time evolution of an effective Hamiltonian. The late-time steady state of a quantum circuit is realized by the thermal equilibrium state or the ground state of the Hamiltonian. Although deriving this Hamiltonian can be challenging, previous research has demonstrated that the effective Hamiltonian can be mapped onto a ferromagnetic Ising model via a replica trick for specific cases of hybrid random circuits \cite{Bao2021, Block2022, Dias2023, Bao2024}. In addition, MIPT observed in these systems corresponds to the transition of two-dimensional bond percolation \cite{fisher_2019,Skinner2019,Yaodong2021_v2}. Thus, the study (i.e., statistical mechanical properties of phase transitions) of MIPT has attracted much attention.

The above-mentioned MIPT occurs for pure states. In cases of initially mixed states, purification phase transition appears instead of the entanglement transition in MIPT, and the nature of the purification phase transition is measured by using logarithmic purity  \cite{Gullans2020}. The measurement purifies the state, and there exists a critical measurement probability for the purification phase transition. 
Although both MIPT and the purification phase transition are induced by measurements, they reflect different effects: measurements break local entanglement in the former, while measurements purify the system in the latter. This difference highlights the distinct roles measurements play in the induction of phase transitions. In this sense, it is interesting to examine the entanglement structure in mixed states.

These two phase transitions are strongly affected by external noise. In general, noise on a circuit arises randomly in space. Especially, we investigate the effects of the spatial non-uniformity in a circuit on purification phase transitions. In this work, we consider its effects in projective measurements and local two-qubit random unitaries, instead of applying a standard noise model like a local quantum channel.
This consideration is qualitatively related to a disordered effective Hamiltonian. In statistical mechanics, the stability of critical points against disorder is evaluated using the Harris criterion \cite{Harris1974}, which focuses on the correlation length critical exponent. In particular, when disorder affects the universality class of a quantum phase transition, the critical behavior can change, potentially altering the value of the correlation length critical exponent itself. Understanding whether this phenomenon occurs even in purification phase transitions is important.
In this work, we employ many-body negativity (MBN) \cite{peres1996,Horodecki_family,weinstein_nega_PhysRevLett.129.080501}, which is a suitable entanglement measure in the random quantum circuit to detect the critical exponent in mixed states. Previous works \cite{sang2021, shi2021} have successfully characterized MIPT by MBN under a subsystem partition, they mainly focus on pure states.
However, the application to mixed states, purification dynamics and its phase transition, as well as detailed numerical observations for them, are still lacking. 
In this work, we numerically investigate the behavior of MBN precisely.
In the steady state characterized by MBN, it sometimes shows different power-law scalings in the structure and model of the system. For example, a model in the random Clifford circuit with boundary decoherence has $L^{1/3}$ power law \cite{weinstein_nega_PhysRevLett.129.080501} (where $L$ is system size). It is curious how MBN scales in our models. In this work, we find that the purification phase transition is captured by MBN, indicating that the phase transition is captured by quantum entanglement quantified in the mixed state. Moreover, we find that the criticality of the purification phase transition is changed in the spatially modulated measurements. This change can be understood by Harris criterion. It suggests that we can also apply its criterion in quantum circuits.
With respect to spatial non-uniformity in local two-qubit random unitaries, we expect distinct localized states from the one observed in the above case to emerge 
because the action of local two-qubit random unitaries on a few sites leads to the formation of short-range entanglement. The short-range entanglement is in contrast to the uniform case described above, where unitary gates act on all sites, leading to the formation of long-range entanglement (excluding the effect of measurements). 
 We consider the quasi-periodic modulation under unitary gates. The modulation could alter the nature of the steady state. We identified a pure-like phase, distinguishable from the pure phase by the entanglement structure of the subsystems. We also find that the effective Hamiltonian for the spatially modulated gates is given by a quasi-modulated ferromagnetic $ZY$ coupling model. These results show that the spatial modulation significantly affects the purification phase transition.

The rest of this paper is organized as follows. 
In Sec.~II, we introduce a random Clifford unitary circuit with on-site projective measurements and a spatial modulation for measurement probability.
In Sec.~III, physical quantities of our interest are explained. There, logarithmic purity and MBN are introduced. In Sec.IV, numerical results are shown. From the behaviors of MBN and logarithmic purity, the transition properties and their quantum criticality are numerically investigated in detail.
In Sec.~V, we provide a discussion of the numerical results obtained in Sec.~IV. In particular, the effects of spatially modulated probability on the criticality from the view point of the Harris criterion are discussed.
In Sec.~VI, we numerically investigate the effect of spatially modulated application of the random Clifford gate with a certain uniform measurement probability for this purification dynamics. 
Section VII is devoted to conclusion.

\section{Random Clifford circuit with spatially modulated measurement probability}
\begin{figure}[t]
\begin{center} 
\vspace{0.5cm}
\includegraphics[width=9cm]{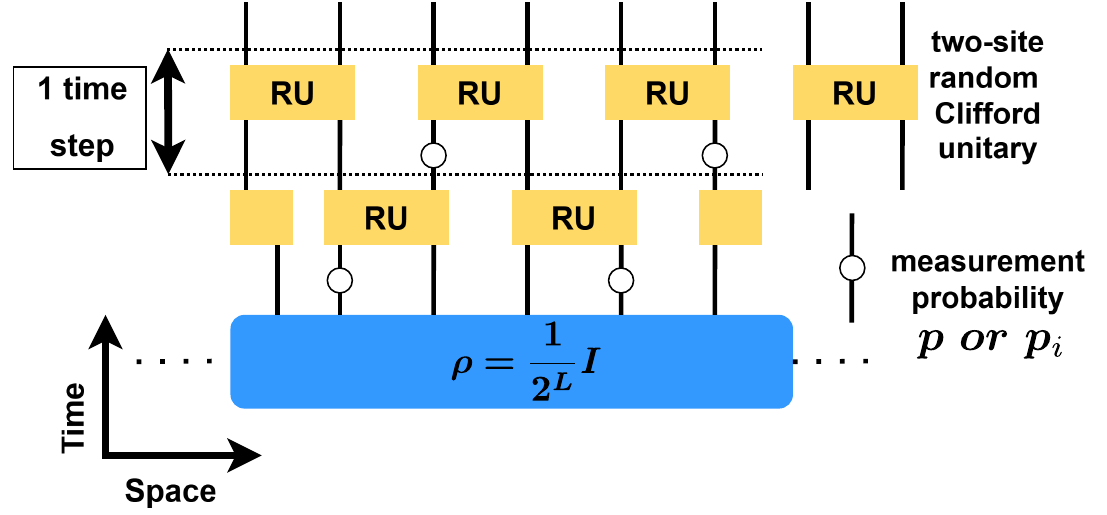} 
\end{center} 
\caption{A schematic illustration of hybrid random Clifford circuit. 
The yellow block represents a two-site random Clifford unitary gate and the small white circle represents a projective measurement of on-site $Z_i$ measurement operator with measurement rate $p$ or $p_i$. One time step consists of one measurement layer and a layer of two-site random Clifford unitaries, where each unitary gates do not overlap.}
\label{Fig1}
\end{figure}

We consider a one-dimensional system of $L$ qubits with periodic boundary conditions and study hybrid random Clifford circuits \cite{Skinner2019,fisher_2019,purification_PhysRevX.10.041020} (see Fig.~\ref{Fig1}). The circuits consist of two parts: the first one is a two-site random unitary layer, where the two-site gate is picked up from the Clifford group \cite{unitary_10.1063/1.4903507,pal_random2site_uni_PhysRevResearch.5.L012031} randomly, and the second one is a measurement layer where each site is measured by a projective measurement of on-site $Z_i$ measurement operator with a uniform measurement probability $p$ or a {\it site-dependent} measurement probability $p_i$, where $i$ is a site number ($i=0,1,\cdots,L-1$). We employ a spatially modulated probability given by $p_i\equiv w_i^n$, where $w_i$ is a uniform random value in $[0,1]$ and $n$ controls the average probability for the entire system given by $\bar{p}=1/(n+1)$ \cite{Zabalo2023}. 
The circuit with $\bar{p}=1/(n+1)$ can be compared with the uniform probability case previously studied \cite{Gullans2020}. One time step is defined as a pair consisting of a unitary layer followed by a measurement layer. In the unitary layer, local two-qubit unitaries act on neighboring pairs of qubits. Each unitary layer has $L/2$ two-qubit unitary gates, acting on all the odd links in the first layer, and all the even links in the second, as illustrated in Fig.~\ref{Fig1}. We set an initial state to be an infinite temperature state $\rho(0)=\frac{1}{2^L}I$.

Throughout this work, we focus on a late-time state. We consider a total number of time steps that scale as $\mathcal{O}(L)$ when analyzing the logarithmic purity \cite{Gullans2020, Ashida2024, Zou2023}, and $\mathcal{O}(L^2)$ when analyzing the MBN.
We find that MBN requires more time steps to saturate than logarithmic purity (see the behavior shown in Appendix B). The previous study \cite{Gullans2020} found that the purification phase transition is in the same universality class as the class of MIPT \cite{Skinner2019,fisher_2019,Li2021}. We are interested in how the purification phase transition is affected by the spatial non-uniformity of the measurement layer. 
We numerically investigate purification phase transitions by employing an efficient numerical algorithm for the stabilizer formalism 
\cite{gottesman1998heisenbergrepresentationquantumcomputers,aaronson_gotttesman_hysRevA.70.052328}, which can carry out large-scale numerical calculations. Technically, two-site random Clifford unitary gates are implemented by a list-up table scheme suggested by \cite{Koenig_Smolin2014,Richter2023}. 

Moreover, in a pure initial state, the late-time state for the random Clifford circuit is related to the ground state obtained by an imaginary time evolution of an effective spin Hamiltonian \cite{Bao2020}. Although this is obtained by approximating a circuit model, it allows us to make qualitative predictions about the late-time state. Based on the relation \cite{Bao2020}, we derive that the circuit with the spatially modulated probability is related to an imaginary time dynamics of the following disordered ZY spin model with random transverse field, 
\begin{eqnarray}
H_{\rm eff}\equiv \sum_{j}\left(-\frac{2}{5}Z_jZ_{j+1}-\frac{1}{10}Y_jY_{j+1}\right)-\sum_{j}\left(\frac{2}{5}+2\gamma_j\right)X_j.\nonumber\\
\label{eHam0}
\end{eqnarray}
A brief derivation of the effective Hamiltonian $H_{\rm eff}$ is shown in Appendix A, from which we find that $\gamma_j \propto p_j$.
The late-time state of the circuit is related to the saturation state of the imaginary-time evolution of the Hamiltonian 
\begin{eqnarray}
\rho(t\to \infty) \approx \lim _{t\to \infty}e^{-t H_{\rm eff}}|\rho_0\rangle\rangle,
\end{eqnarray}
where $|\rho_0\rangle\rangle$ represents a supervector state of $(\rho_0)^{\otimes 2}$. It consists of two copies of the initial density matrix.
From this picture, the random transverse field in the Hamiltonian $H_{\rm eff}$ gives an impact on the criticality of the original Hamiltonian before introducing the random transverse field. This influence resembles disordered quantum phase transitions, as discussed in \cite{sachdev2011quantum, chandran_PhysRevX.7.031061}. With the help of the view points of the effective Hamiltonian and its ground states, we numerically investigate the late-time state.

\section{Target physical quantities}
We explain the target physical quantities of this work.
The first one is the logarithmic purity denoted as $S_P$ \cite{Gullans2020}, 
\begin{eqnarray}
S_P=-\log \Tr[\rho^2].
\label{purity_eq}
\end{eqnarray}
This is also called the second-order R\'{e}nyi entropy for the density matrix $\rho$ \cite{Ashida2024,Zou2023}.
The value is efficiently calculated in the stabilizer formalism \cite{gottesman1998heisenbergrepresentationquantumcomputers, aaronson_gotttesman_hysRevA.70.052328}.
Practically, it is calculated from the Gaussian elimination \cite{sang2021}.

The second one is the many-body negativity(MBN), defined as 
$$
\mathcal{E}_A\equiv \log_2|\rho^{\Gamma_A}|_1.
$$
where $A$ is a half-chain subsystem, $|\cdot|_1$ represents the trace norm, and $\Gamma_A$ is a partial transpose operation for A-subsystem. 
This quantity is a many-body version of the one for a few-spin system firstly proposed in \cite{peres1996,Horodecki_family,log_nega_PhysRevA.65.032314}. 
The negativity can reveal a transition of states by quantifying quantum entanglement in mixed states \cite{lu2020,sang2021,weinstein_nega_PhysRevLett.129.080501}. 
We explain the calculation of MBN \cite{sang2021,shi2021,sharma2022,KOI2023_1} (the detailed derivation is described in Refs.~\cite{sang2021,shi2021}). 
As shown in Refs.~\cite{sang2021,shi2021,sharma2022,KOI2023_1}, MBN is calculated by a following formula 
\begin{eqnarray}
\mathcal{E}_{A}=\frac{1}{2}\mathrm{rank}J,\nonumber
\end{eqnarray}
where $J$ is a $m \times m$ matrix, $m$ is the total number of the stabilizer generator for a state $\rho$ and here $m \le L$.
The matrix $J$ is constructed by the anticommutation for a truncated $m$-stabilizer generators $g^A_{n}$ ($n=0,\cdots, m-1$) of the state $\rho$. The truncated stabilizer generators are created as follows: we represent the stabilizer generators as the binary representation denoted by $g_{n}$ \cite{Nielsen2011}. 
Then, we truncate the binary representation vectors of each stabilizer generator to remove the element of sites not in the subsystem A as 
\begin{eqnarray}
g_{n} \longrightarrow g^A_{n}=(g^{n,x}_0,\cdots, g^{n,x}_{k} | g^{n,z}_0,\cdots, g^{n,z}_{k}),\nonumber
\end{eqnarray} 
where $g^{n,x(z)}_{p}=0$ or $1$ and the binary components with the qubit label of the subsystem $A$ (here labeled by $0,\cdots, k$) remain.  
Finally, by using the $m$ truncated stabilizer generators $g^A_{n}$, we construct the matrix $J$ given by 
\begin{eqnarray}
(J)_{n,n'}=
\begin{cases}
1 & \mbox{if}\:\: \{g^{A}_{n},g^{A}_{n'}\}=0\\
0 & \mbox{if}\:\: [ g^{A}_{n},g^{A}_{n'}]=0
\end{cases}
,\nonumber
\end{eqnarray}
where $\{g^{A}_{n},g^{A}_{n'}\}=0$ means that the truncated stabilizer generators $g^{A}_{n}$ and $g^{A}_{n'}$ are anti-commuting,
and $[g^{A}_{n},g^{A}_{n'}]=0$ means that the truncated stabilizer generators $g^{A}_{n}$ and $g^{A}_{n'}$ are commuting.
Thus, the matrix $J$ is the binary $m \times m$ matrix. Finally, the calculation of the rank of $J$ gives the value of MBN.

we consider the mean value of the physical quantities given by
\begin{eqnarray}
\langle Q\rangle=\mathbb{E}[Q^s],\nonumber
\end{eqnarray} 
where $Q=S_P$ or $\mathcal{E}_A$, $\langle\cdot\rangle$ represents the mean value, and $Q^s$ represents a single sample (labeled by $s$) of $Q$.

\section{Numerical results}
In this section, we show numerical results. In particular, we investigate late-time states almost saturated after $\mathcal{O}(L)$ time steps for the logarithmic purity and after $\mathcal{O}(L^2)$ time steps for MBN.

\begin{figure}[t]
\begin{center} 
\vspace{0.5cm}
\includegraphics[width=7.5cm]{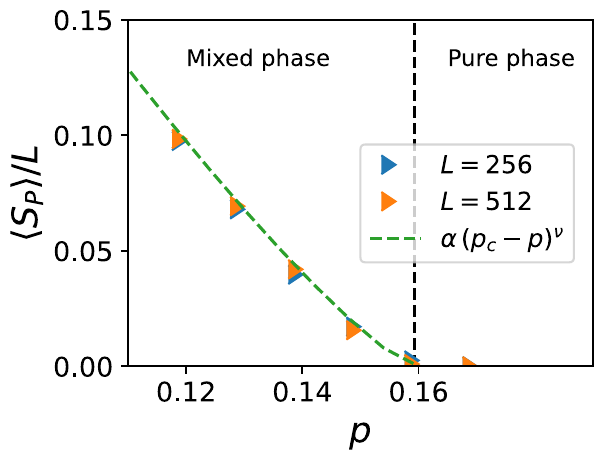}  
\end{center} 
\caption{
Logarithmic purity with uniform measurement probability $p$, plotted as a function of $p$ for various system sizes $L$.
We took total time steps $t_{step}=4L$.
The green curve shows $\alpha(p_c-p)^\nu$ for $p \le p_c=0.159$ and $\nu= 1.22$.}
\label{Fig2}
\end{figure}
\subsection{Case I: Uniform measurement probability $p$}
We first show the uniform measurement probability case. That is, all sites on the measurement layer can be measured with a probability $p$, which is the same setup in \cite{Gullans2020}. 

\begin{figure}[t]
\begin{center} 
\vspace{0.5cm}
\includegraphics[width=8.5cm]{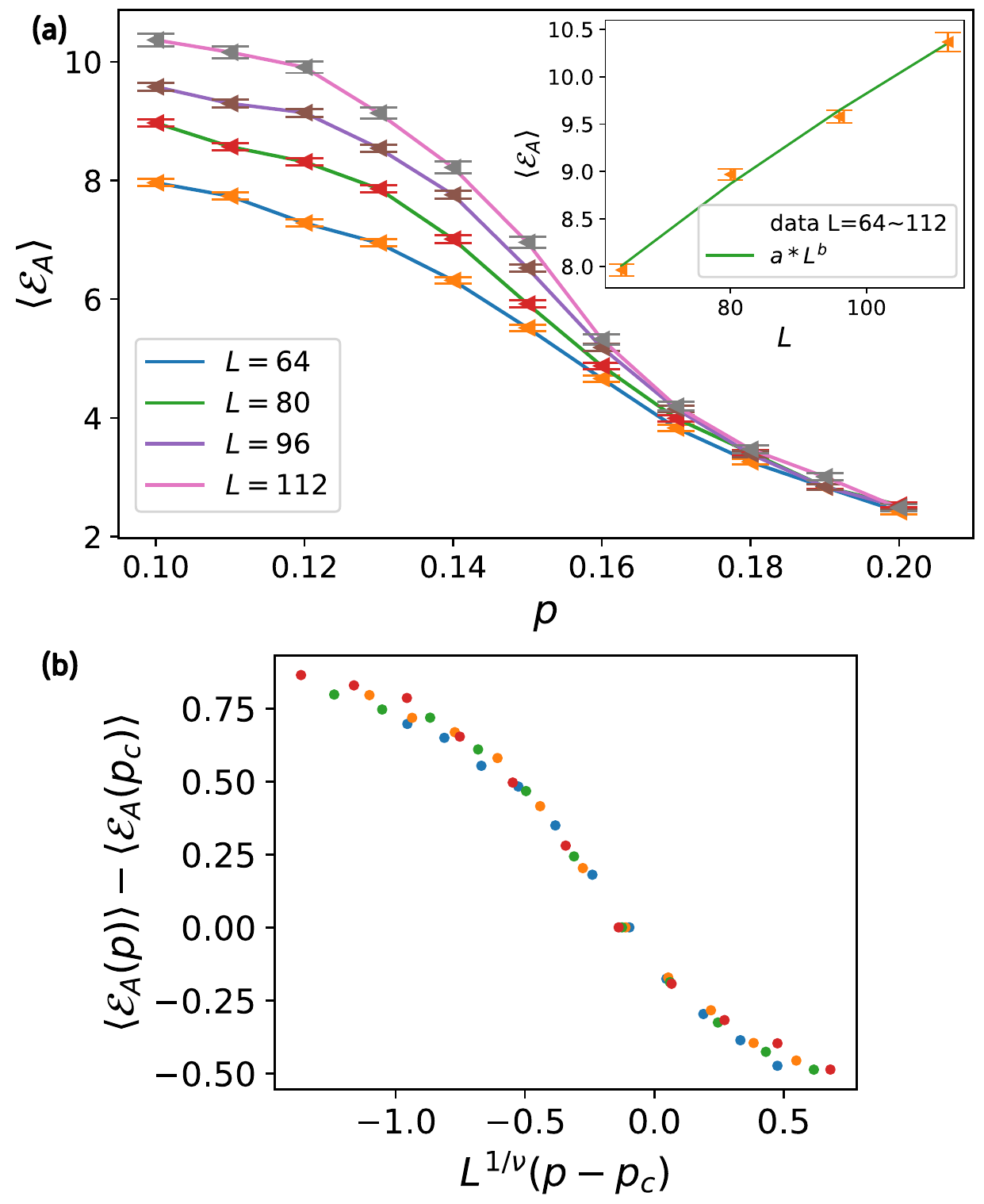}  
\end{center} 
\caption{(a)MBN with uniform measurement probability $p$, plotted as a function of $p$ for various system sizes $L$.
We took total time steps $t_{step}={L^2}/{4}$.
The inset represents the scaling behavior of MBN at a typical mixed phase $p=0.10$, where the fitting function is set as $y=aL^b$ (green line) and we extracted $a \approx 1.18$ and $b \approx 0.460$. (b) Scaling data collapse for different system size data of MBN. By using $L=64\sim112$
 data, we obtain $p_c=0.167\pm0.00118,\, \nu=1.56\pm0.108$. We averaged $\mathcal{O}(10^2)$ samples for each data point.}
\label{Fig3_clean_negativity}
\end{figure}
In Fig~\ref{Fig2}, we show the logarithmic purity with uniform measurement probability $p$, plotted as a function of $p$ for various system sizes $L$. We reproduced the same behavior of the logarithmic purity to that of \cite{Gullans2020}.  Two distinct phases appear for $p < 0.16$ and $p > 0.16$, respectively: mixed and pure phases. A purification phase transition exists between these phases. In the mixed phase for $p < 0.16$, we observed that $\langle S_P \rangle \approx L$, indicating a volume law scaling. In contrast, in the pure phase for $p > 0.16$, $\langle S_P \rangle \approx 0$. Its behavior is independent of the system size, corresponding to the area law scaling. These size scalings are same to the volume-law and area-law phases known in the study of MIPT \cite{fisher_2019,Skinner2019}.
We employ a scaling form $\langle{S_{P}}\rangle/L\approx\alpha(p-p_c)^\nu$ ($\alpha$ is a constant) for $p<p_c$ to confirm the previous results $p_c=0.1593,\nu=1.28$ in \cite{Gullans2020}. We find $p_c=0.159$, $\nu=1.22$. These values are almost consistent with the previous study.

We next observe MBN which is a main focus in this study. This quantity can extract quantum entanglement by partitioning the system in a mixed state. 
Numerical results of MBN as a function of $p$ are displayed in Fig.~\ref{Fig3_clean_negativity}(a).
We observed a system-size dependence in $p<0.17$. 
On the other hand, in $p>0.17$, we observed area-law scaling ($\langle\mathcal{E}_A\rangle\approx\mathcal{O}(L^0)$).
These numerical results suggest the presence of two different phases bordering on $p\approx0.17$: mixed phase and pure phase. 
We evaluated a system-size dependence of MBN at $p=0.1$ by a scaling form $\langle \mathcal{E}_A(p=0.1,L)\rangle=aL^b$ (see the inset of Fig.~\ref{Fig3_clean_negativity}(a)). We extracted $a\approx 1.18$ and $b\approx0.460$.

Furthermore, we elucidate the phase transition point and its criticality in terms of MBN. 
Accordingly, we employ the following ansatz. 
\begin{eqnarray}
\langle\mathcal{E}_A(p)\rangle-\langle\mathcal{E}_A(p_c)\rangle=L^{\zeta/\nu}\tilde{f}\left(L^{1/\nu}(p-p_c)\right),
\label{nega_scaling_formula}
\end{eqnarray}
where $p_c$ is a critical transition probability, $\zeta$ is a critical exponent and $\nu$ is a correlation length critical exponent.
Eq.~(\ref{nega_scaling_formula}) is a more general ansatz than one employed in \cite{weinstein2024} because of the presence of $\zeta$.
By making use of the python package pyfssa \cite{melchert2009,pyfssa}, 
the finite-size scaling analysis is carried out. We obtain a clear data collapse as shown in Fig.~\ref{Fig3_clean_negativity}
(b). We find $p_c=0.167\pm0.00118$ and $\nu=1.56\pm0.108$. 
In MIPT with pure initial state \cite{Skinner2019,fisher_2019,Zabalo2020,Li2021}, critical measurement probability and correlation length critical exponent are $p_c=0.16, \nu=4/3$ respectively. Especially with respect to $\nu$, the value of MBN is greater than that of MIPT ($1.56 \pm 0.108>4/3$).
On the other hand, the value of logarithmic purity is smaller than them ($1.22 < 4/3$).
We expect that these values approach $4/3$ as the system size increases.

\subsection{Case II: Spatially modulated probability $p$}
\begin{figure}[t]
\begin{center} 
\vspace{0.5cm}
\includegraphics[width=6.5cm]{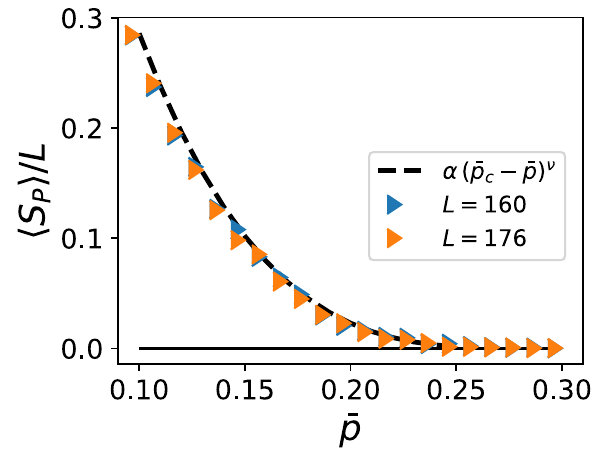}  
\end{center} 
\caption{Logarithmic purity with averaged probability $\bar{p}$, plotted as a function of $\bar{p}$ for various system sizes $L$.
We took total time steps $t_{step}=4L$. The blue curve shows $\alpha(\bar{p}_c-\bar{p})^{\nu}$ for $\bar{p}\le\bar{p}_c=0.290$ and $\nu=3.38$. Each data point is calculated by using $\mathcal{O}(10^2)$ samples.}
\label{Fig4}
\end{figure}
We next observe the spatially modulated probability case. The measurement probability is site dependent, with each site $j$ having its own measurement probability $p_j$. The initial state, unitary layer and total time steps [A typical time evolution is shown in Appendix B] are same as those of Case I. 

In Fig~\ref{Fig4}, we show the logarithmic purity, plotted as a function of the average measurement probability $\bar{p}$ for various system sizes $L$.
A purification phase transition is also observed around $\bar{p} \approx 0.29$.
Compared with Fig.~\ref{Fig2}, we can see a shift of the transition point $p_c=0.159$ for Case I to $\bar{p}_c=0.290$ for Case II.
From the behavior of the logarithmic purity, we elucidate the transition by adopting a scaling form $\langle{S_{P}}\rangle/L\approx\alpha(\bar{p}-\bar{p}_c)^{\nu}$ with the numerical results, where $\bar{p}_c$ and $\nu$ are the critical transition probability and the correlation length critical exponent from the point of view of the logarithmic purity. We evaluated $\bar{p}_c=0.290$ and $\nu=3.38$ respectively. These results are explicitly different from uniform measurement probability (Case I) \cite{Gullans2020}.

\begin{figure}[t]
\begin{center} 
\vspace{0.5cm}
\includegraphics[width=9cm]{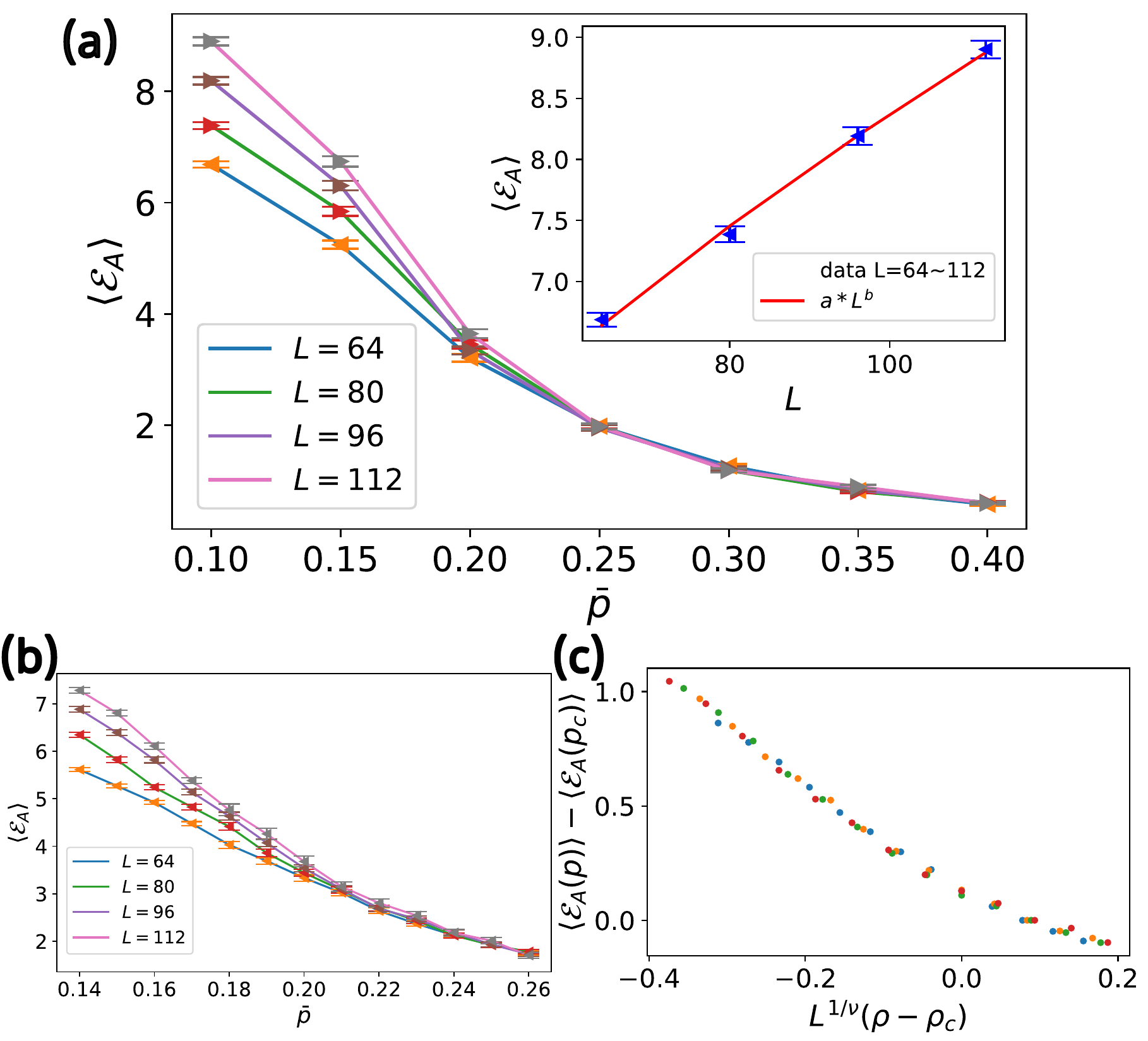}  
\end{center} 
\caption{(a)
MBN with the averaged measurement probability, plotted as a function of $\bar{p}$ for various system sizes $L$.
We took total time steps $t_{step}={L^2}/{4}$.
The inset represents the scaling behavior of MBN at a typical mixed phase $\bar{p}=0.10$, where the red line is fitting curve $y=aL^b$  with $a \approx 0.767$ and $b \approx 0.519$. 
(b) The details of the data of MBN around the purification phase transition point. 
(c) Scaling data collapse for different system size data of MBN. By using $L=64\sim112$
 data, we obtain $\bar{p}_c=0.220\pm0.00261,\, \nu=3.06\pm0.313$. We averaged $\mathcal{O}(10^3)$ samples for each data point.}
\label{Fig5}
\end{figure}

We observe MBN for this spatially modulated probability case. In Fig.~\ref{Fig5}(a), MBN is plotted as a function of the averaged measurement probability $\bar{p}$ for various system sizes $L$.
The results indicate that MBN exhibits a system-size dependence for $\bar{p}<0.20$. Additionally, it obeys the area-law for $\bar{p}>0.20$. These behaviors indicate the presence of two different phases: mixed and pure phase. For $\bar{p}<0.20$, MBN exhibits a mixed phase.
We also evaluate a system-size dependence of MBN at $\bar{p}=0.1$ by a scaling form $\langle \mathcal{E}_A(\bar{p}=0.1,L)\rangle=aL^b$. We extract $a\approx 0.767$ and $b\approx0.519$. 
We estimate the phase transition point and its criticality in terms of MBN. We use the same finite-size scaling ansatz [\ref{nega_scaling_formula}] to that of Case I. The finite-size scaling analysis indicates a clear data collapse as shown in Fig.~\ref{Fig5}
(c). We extract $\bar{p}_c=0.220\pm0.00261$ and $\nu=3.06\pm0.313$.

From these numerical estimations, we observed a difference in the correlation length critical exponent between Case I and Case II.

\section{Comparison and Discussion of Numerical Results for Case I and Case II}
In this section, we focus on two main points. First, in both Case I and Case II, we discuss the results that MBN and logarithmic purity exhibit distinct power-law scaling behaviors with system size in the mixed phase. Second, the difference in the correlation length critical exponents between Case I and Case II can be understood in terms of the Harris criterion \cite{Harris1974}.

Regarding the first point, we performed a size scaling analysis of the MBN at measurement rate $p = 0.1$, assuming the fitting function $\langle \mathcal{E}_A(p=0.1, L) \rangle = a L^b$.
As a result, we numerically extracted the size exponent $b = 0.460$ for Case I and $b = 0.519$ for Case II. In contrast, the logarithmic purity $\langle S_P \rangle$ scaled as $\langle S_P \rangle / L \approx \alpha ({p} - {p}_c)^{\nu}$ for both Case I and Case II, indicating that $\langle S_P \rangle$ is proportional to the system size $L$. 
The behavior of $\langle S_P \rangle$ is different from that of MBN because both $b$ of Case I and Case II are less than $1$. This is because MBN is able to capture only quantum correlations. According to \cite{log_nega_PhysRevA.65.032314}, negativity can filter out such classical correlations in a few-spin systems even though mixed states include classical correlations.
We expect that many-body negativity (MBN) excludes classical correlations even in many-body settings.
Furthermore, we compare our results with a previous study using the MBN~\cite{weinstein_nega_PhysRevLett.129.080501}, in which a local dephasing was applied to the boundaries. In that setting, it is known that the MBN scales as the $1/3$ power of the system size. This value is smaller than both of our calculated exponents, $b = 0.460$ and $b = 0.519$.

Regarding the second point, we explain Harris criterion at first. After that, we discuss our numerical results by Harris criterion. 
For classical systems, Harris showed a phenomenological argument of the correlation length critical exponent $\nu$ \cite{Harris1974}, namely Harris criterion. 
The criterion is also applicable to quantum phase transitions in $(d+1)$-dimensional space-time system (where $d$ is a spatial dimension) \cite{Chayes_criterion_PhysRevLett.57.2999,sachdev2011quantum}. Harris criterion in the version of quantum phase transitions claims that the phase transition in systems without disorder is not affected by disorders if ${\nu \ge {2}/{d}}$.
This argument by Harris can also be interpreted as follows: If there exists a quantum phase transition in the system without disorder with ${\nu < {2}/{d}}$, a disorder can change the universality class of the phase transition. This interpretation and application of the Harris criterion are found in \cite{Peng2020}. 

We discuss our numerical results based on the Harris criterion.
The quantum circuits we consider are in one spatial dimension, hence $d=1$. In Case I, we obtained correlation length critical exponents of $\nu=1.22$ for $S_P$ and $\nu=1.56\pm0.108$ for $\mathcal{E}_A$, respectively. Remarkably, both of $\nu$ are less than 2. According to the Harris criterion, for the quantum circuit model in Case I, if a phase transition persists after adding perturbations, its universality class is expected to change. In fact, the values of $\nu$ we obtained for Case II are $\nu=3.38$ for $S_P$ and $\nu=3.06\pm 0.313$ for $\mathcal{E}_A$, respectively. Both $S_P$ and $\mathcal{E}_A$ of $\nu$ are different from Case I. Therefore, Case II maintains the purification phase transition and furthermore alters the universality class of quantum criticality. Our result obtained in the quantum circuit is in agreement with the Harris criterion.

\section{Spatially modulated probability for random Clifford unitary gate}
In this section, we investigate the effect of spatial non-uniformity of random Clifford unitary gate. This case also provides further insight into understanding the effects of spatial non-uniformity to the circuit.

We introduce the modulation for two-site random Clifford unitary gate for each nearest-neighbor site. 
Here, we consider a spatially modulated probability $p_j^u$\,($j$ means the link between $j$-th and $j+1$-th sites) defined by the following formula  
\begin{eqnarray}
p^{u}_j = \frac{1+A_J\cos(2\pi Q j)}{1+A_J},
\end{eqnarray}
where $0\leq A_J \leq 1$ and $Q=\frac{1+\sqrt{5}}{2}$ (golden ratio). 
We then fix the uniform measurement probability at $p=0.1$ for the measurement layer. 

In the effective Hamiltonian picture, the effect of $p^{u}_j$ can be regarded as spatially modulated couplings in the $ZY$ term. 
The effective Hamiltonian is
\begin{eqnarray}
H_{\rm eff} \equiv \sum_{j}
    & \tilde{J}_{j} \left( -\frac{2}{5} Z_j Z_{j+1} - \frac{1}{10} Y_j Y_{j+1} \right) \notag \\
    &+ \displaystyle\sum_{j} \left( -\frac{2}{5} - 2\gamma_0 \right) X_j ,
\label{eHam2}
\end{eqnarray}
where $\tilde{J_{j}}\propto p^u_j$, $\gamma_0 \propto p=0.1$.
This spatial modulation acts like a quasi-periodic ferromagnetic $ZY$-coupling. 
As getting large $A_J$,  entanglement generation is suppressed due to the reduced probability of unitary gate application.
From the effective Hamiltonian picture, we expect that even in the circuit system, the effects of the modulation give a change of the purification dynamics and its late-time state and there is a critical $A_J\,(p_j^u)$ to change entanglement structure.

\begin{figure}[t]
\begin{center} 
\vspace{0.5cm}
\includegraphics[width=8.5cm]{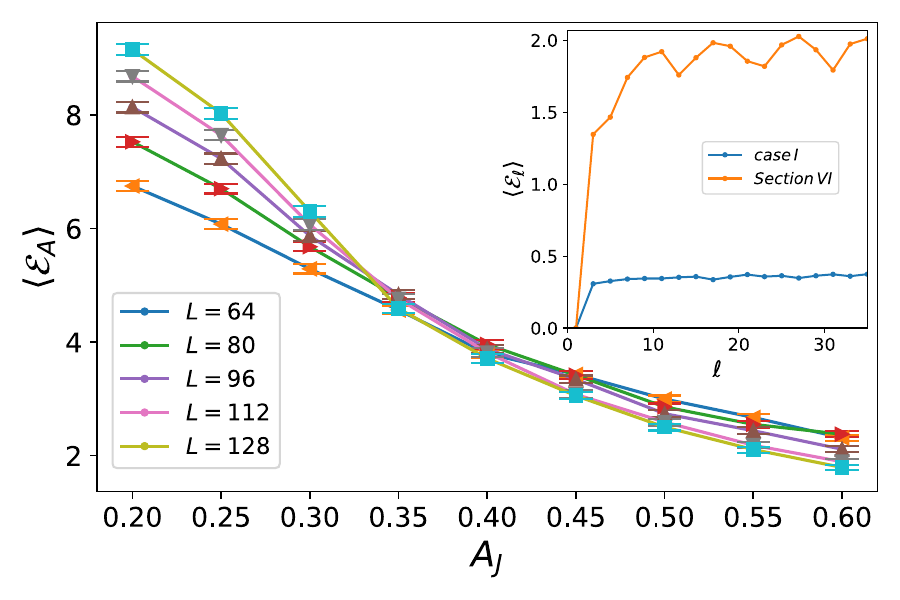}  
\end{center} 
\caption{$A_J$-dependence of MBN for the late-time state at $t_{step}={L^2}/{4}$. We averaged $\mathcal{O}(10^3)$ samples for each data point. The inset represents the subsystem-size dependence of MBN at a mixed phase for $A_J=0.6$, $p=0.1$, $L=96$ (the orange line). As a comparison, we plot a uniform probability ($p=0.4$) case (Case I) shown as the blue line.}
\label{Fig6}
\end{figure}

In this setting of unitary gates, we numerically observe the MBN of the late-time state as varying $A_J$. 
The results are shown in Fig.~\ref{Fig6}.
A data crossing appears at $A_J\approx 0.35$. This data crossing implies the presence of a phase transition. 
In the region where  $A_J<0.35$, a mixed phase is observed in which the value of MBN increases as the system size increases. In contrast, for $A_J>0.35$, MBN exhibits a tiny decrease as the system size increases. Hereafter, for convenience, we refer to the regime of $A_J>0.35$ as the pure-like phase. This pure-like phase is different from the pure phase of both Case I and Case II shown in the previous section, where MBN exhibits system-size independence (area law scaling).

In the inset of Fig.~\ref{Fig6}, to elucidate the entanglement structure and its difference from that of the previous area law phases in Case I and Case II, we observe the subsystem-size dependence of MBN where the subsystem size denoted by $\ell$. The MBN is given by
$
\mathcal{E}_\ell\equiv \log_2|\rho^{\Gamma_\ell}|_1.
$
The mean value of $\mathcal{E}_\ell$, denoted as $\langle \mathcal{E}_\ell\rangle$, is obtained by averaging over different samples. The orange line in the inset of Fig.~\ref{Fig6} represents $\langle \mathcal{E}_\ell\rangle$ in the pure-like phase regime, with parameters $A_J=0.6$, $p=0.1$, and $L=96$ respectively.
We show the subsystem-size dependence of MBN in the pure phase with $p=0.4$ in Case I as a comparison, the result of which is the blue line in the inset of Fig.~\ref{Fig6}.
The result indicates that MBN for the quasi-periodic unitary case is larger than that of Case I. It implies the pure-like phase exhibits short-range entanglement. Then, we observe an oscillating behavior of MBN in this model as varying $\ell$, the period of oscillation corresponds to the period of the quasi-periodic modulation probability. This indicates that the subsystem-size dependence of the entanglement is strongly affected by the quasi-periodic modulation of two-site unitary gate.

\section{Conclusions}
We numerically investigated the effect of spatial modulation in a random Clifford circuit and its mixed-state dynamics by focusing on MBN. We found that MBN is valid to extract internal quantum entanglement for mixed states and numerically demonstrated that MBN is a good indicator to characterize the purification phase transition and to
detect its criticality, especially the correlation length critical exponent $\nu$ and critical measurement probability $p_c$. In Case I, we found $p_c=0.167\pm0.00118,\, \nu=1.56\pm0.108$. On the other hand, in Case II, we found that the spatially modulated probability gives significant effects on the criticality of the purification phase transition as $\bar{p}_c=0.220\pm0.00261,\, \nu=3.06\pm0.313$. The main result of our study is that MBN captures the changes in the correlation length critical exponent $\nu$ in Case I and Case II.
This change of the quantum criticality is understood by using Harris criterion. 
In Case I, the correlation length critical exponent satisfies $\nu=1.22<2$ for $S_P$ and $\nu=1.56\pm0.108<2$ for $\mathcal{E}_A$, respectively (where $d=1$). Through Harris criterion, a disorder changes the universality class of the phase transition.
In fact, in Case II, the correlation length critical exponent in both the logarithmic purity and MBN differs from Case I, suggesting a distinct universality class.
These results are summarized in the following TABLE. 
\begin{table}[h]
    \centering
    \label{result_summary}
    \begin{tabular}{|c|c|c|}
        \hline
        &the logarithmic purity&MBN\\
        \hline
        Case I & $\nu=1.22$ & $\nu=1.56\pm0.108$ \\ \hline
        Case II & $\nu=3.38$ & $\nu=3.06\pm 0.313$\\ 
        \hline
    \end{tabular}
    \caption{Numerically extracted correlation length critical exponent of Case I and Case II. In this work, we newly estimated the correlation length critical exponent $\nu$ from the MBN in Case I and II, as well as from the logarithmic purity in Case II.}
\end{table}

We conclude that the spatial modulation in the measurement probability gives the change of the universality class of the purification phase transition from the viewpoint of the MBN. 
Furthermore, the change in the correlation length critical exponent is consistent with the insight discussed in Section II, Eq.~(\ref{eHam0}), namely that the random transverse field affects quantum criticality.

Moreover, we numerically investigate the effect of introducing a quasi-periodic spatially modulated probability for random Clifford unitary gates. This modulation also induces a purification phase transition, but the pure-like phase appears to differ from the pure phase in Cases I and Case II in terms of the subsystem dependence of MBN. It indicates that a unique internal correlation pattern appears in the pure-like phase. According to \cite{chandran_PhysRevX.7.031061,quasi_PhysRevB.107.134201}, it is known that quasiperiodic modulation induces a quantum phase transition. Therefore, the difference between the pure-like phase and the pure phase can be understood as the difference of localized phase.

Apart from numerical analysis of the circuit, some master equations can describe the circuit dynamics \cite{zhou_10.21468/SciPostPhysCore.6.1.023}.
It is an interesting future subject to understand measurement-induced (purification) phase transitions from the theoretical flame work of the master equations and to further investigate the relationship between numerical results and these equations.

\section*{ACKNOWLEDGMENTS}
This research was supported in part by JSPS KAKENHI Grant Numbers JP24K06878, JP24K00563, JP24K02948, JP21H04446, and CSIS in Tohoku University.

\renewcommand{\thesection}{A\arabic{section}} 
\renewcommand{\theequation}{A\arabic{equation}}
\renewcommand{\thefigure}{A\arabic{figure}}
\setcounter{equation}{0}
\setcounter{figure}{0}

\appendix
\section*{Appendix A: Effective spin model}
In this Appendix, we briefly show the derivation of the effective Hamiltonian shown in the main text [Eq.~(1)] by reviewing the analytical manipulations in Refs.\cite{Bao2021,Block2022,Dias2023,Bao2024}. 
We show that dynamics of a hybrid random quantum circuit can be related to an imaginary time dynamics with an effective quantum spin model.

\begin{figure}[t]
\begin{center} 
\vspace{0.5cm}
\includegraphics[width=5.5cm]{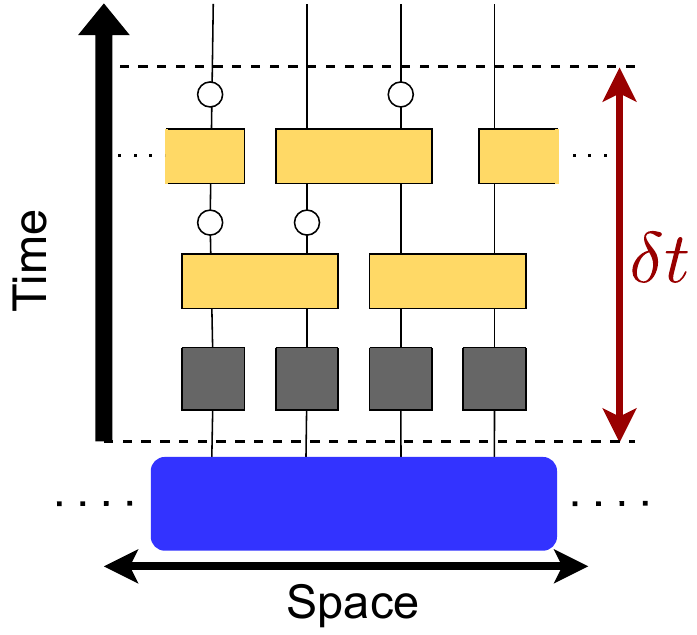}  
\end{center} 
\caption{
Hybrid (Haar) random circuit. 
The yellow block represents a two-site Haar random unitary gate, the gray block represents single-site Haar random unitary gate and the small white circle represents a projective measurement with on-site $Z_j$ measurement operator where the measurement rate is site-dependent $p_i$.
This model can be qualitatively related to the circuit model in the main text(shown in Fig.~\ref{Fig1}).}
\label{FigA1}
\end{figure}

As a related circuit to the one in the main text (shown in Fig.~\ref{Fig1}), we consider a Haar random circuit shown in Fig.~\ref{FigA1}, where the small time-interval includes local a layer of on-site Haar-random unitary applying the every site, two-site random unitary layer including even site pair, two-site random unitary layer including odd site pair and two measurement layers. In the measurement layer, we consider that each site is measured with a spatially modulated probability $p_i$, which is a only difference from the original setup \cite{Bao2021,Block2022,Bao2024}.
The schematic of the small time-interval is shown in Fig.~\ref{FigA1}. 

We first start to two copies of a density matrix labeled by $s$,  $(\rho_s)^{\otimes 2}$. Here, we introduce the supervector representation, $(\rho_s)^{\otimes 2}\longrightarrow |\rho_s\rangle\rangle$ \cite{Bao2024}.
Then, the averaged density matrix over samples (labeled by $s$) is given by \cite{Block2022} 
\begin{eqnarray}
|\rho\rangle\rangle\equiv \sum_{s}|\rho_s\rangle\rangle.
\end{eqnarray}
Note that this vector is unnormalized (the norm is related to the purity of the system). 
We expect that the averaged late-time state for $|\rho\rangle\rangle$ gives qualitative insight to understand the late-time state of the Clifford circuit in the main text. 

Based on the averaged picture, we start to consider the evolution of the small-time interval $\delta t$. The operation is given by \cite{Block2022}
\begin{eqnarray}
&&|\rho(t+\delta t)\rangle\rangle=(MU_{NN_o}MU_{NN_e})U_{1}|\rho(t)\rangle\rangle\nonumber\\
&&=(PMP)\cdot(PU_{NN_o}P)\cdot(PMP)\cdot (PU_{NN_e}P)|\rho(t)\rangle\rangle, \nonumber\\
\end{eqnarray}
where $U_1$ is a layer constituted by the product of single site Haar random unitary (the black squares in Fig.~\ref{FigA1}), $U_{NN_{e(o)}}$ represents unitary layer constituted by two-site Haar random unitaries for even(odd)-links, and
$P$ is the averaged single-qubit Haar random gate, which can be represented by an effective spin-$1/2$ Hilbert space \cite{Block2022},
\begin{eqnarray}
P\equiv \bigotimes^{L-1}_{j=0}\sum_{\sigma=\uparrow, \downarrow}|\sigma\rangle\rangle \langle \langle \sigma|_j.
\end{eqnarray}
Here, $\{|\uparrow\rangle\rangle_j,|\downarrow\rangle\rangle_j\}$ can be regarded as spin-1/2 orthogonal basis on two dimensional Hilbert space on the site $j$ and $P$ acts as a projector \cite{Bao2024}. 

Based on the projector $P$ and employing the transfer matrix picture \cite{Bao2024}, the average action of the two-site Haar random gate can be described in terms of the effective spin $1/2$ bases as 
\begin{eqnarray}
&&PU_{NN_{e}}P\equiv \prod^{L/2}_{\ell=0}\biggl[I+\delta tP\overline{u_{2\ell,2\ell+1}}P\biggr].\:\:
\end{eqnarray} 
By consider an exponent form for the above  we can define a local effective Hamiltonian
\begin{eqnarray}
\mathcal{P}U_{NN}\mathcal{P}\equiv e^{-\delta t \sum_j h^{j,j+1}_{NN}}.
\end{eqnarray}
Here, we omitted the even or odd label of the unitary gates ($U_{NN_{e(o)}}\to U_{NN}$).

We next describe the concrete form of the $h^{j,j+1}_{NN}$. 
The local averaged two-site Haar random gate is given by \cite{Bao2024}
\begin{eqnarray}
&&\overline{u_{j,j+1}}=\overline{(U_{j,j+1}\times U^*_{j,j+1})^{\otimes 2}}\nonumber\\
&=&\frac{1}{15}|\mathcal{I}_j\mathcal{I}_{j+1}\rangle\rangle \langle\langle \mathcal{I}_j\mathcal{I}_{j+1}|+\frac{1}{15}|\mathcal{C}_j\mathcal{C}_{j+1}\rangle\rangle \langle\langle \mathcal{C}_j\mathcal{C}_{j+1}|\nonumber\\
&&
-\frac{1}{60}|\mathcal{I}_j\mathcal{I}_{j+1}\rangle\rangle \langle\langle \mathcal{C}_j\mathcal{C}_{j+1}|-\frac{1}{60}|\mathcal{C}_j\mathcal{C}_{j+1}\rangle\rangle \langle\langle \mathcal{I}_j\mathcal{I}_{j+1}|,\nonumber\\
\end{eqnarray}
where 
\begin{eqnarray}
&&|\mathcal{I}_j\rangle\rangle=c_+|\uparrow\rangle\rangle_j +c_-|\downarrow\rangle\rangle_j,\\
&&|\mathcal{C}_j\rangle\rangle=c_-|\uparrow\rangle\rangle_j +c_+|\downarrow\rangle\rangle_j,
\end{eqnarray}
with $c_\pm=(\sqrt{3}\pm 1)/(\sqrt{2})$.

By using (A4), (A5), (A6), (A7) and (A8), 
$h^{j,j+1}_{NN}$ is given by
\begin{eqnarray}
h^{j,j+1}_{NN}&=&\sum_{\alpha,\beta,\gamma,\lambda=\uparrow, \downarrow}C^{\alpha,\beta}_{\gamma,\lambda}|\alpha\beta\rangle\rangle_{j,j+1}\langle\langle \gamma \lambda|_{j,j+1},\\
C^{\alpha,\beta}_{\gamma,\lambda}&=&{}_{j,j+1}\langle\langle \alpha \beta|\overline{u_{j,j+1}}|\gamma\lambda\rangle\rangle_{j,j+1}
\end{eqnarray}
By representing $4\times 4$ matrix $C^{\alpha,\beta}_{\gamma,\lambda}$ in terms of two pauli matrix product, $h^{j,j+1}_{NN}$ becomes 
\begin{eqnarray}
h^{j,j+1}_{NN}&=&-\frac{2}{5}Z_jZ_{j+1}-\frac{1}{10}Y_jY_{j+1}+\frac{1}{5}(X_j+X_j+1).\nonumber\\
\end{eqnarray}
Here, we dropped a constant identity.

Furthermore, we use the same procedure for the measurement layer $(PMP)$. 
The operator $M$ \cite{Block2022} is given as
\begin{eqnarray}
M&\equiv& \prod_j M_j,\nonumber\\
M_j&=&(1-p_j){\bf 1}^{\otimes 4}+p_j\sum_{\mu=\pm}P^{\otimes 4}_{j,\mu},\:\:P_{j,\mu}=\frac{1+\mu \sigma^z_j}{2}.\nonumber
\end{eqnarray}
Then, the measurement layer $(PMP)$ can be exponetiated as \cite{Block2022}
\begin{eqnarray}
(PMP)&\equiv& e^{-\delta t \sum_{j}h^{m}_j},\:\:
h^{m}_j= \frac{\gamma_j}{3}X_j +(\frac{1}{3}-\gamma_j)I_j,\nonumber\\
\end{eqnarray}
where we set $p_j=\gamma_j\delta t$, the term $h^m_j$ is a {\it spatially modulated local transverse field}. The final identity term in $h^m_j$ can be dropped.

\begin{figure}[b]
\begin{center} 
\vspace{0.5cm}
\includegraphics[width=7cm]{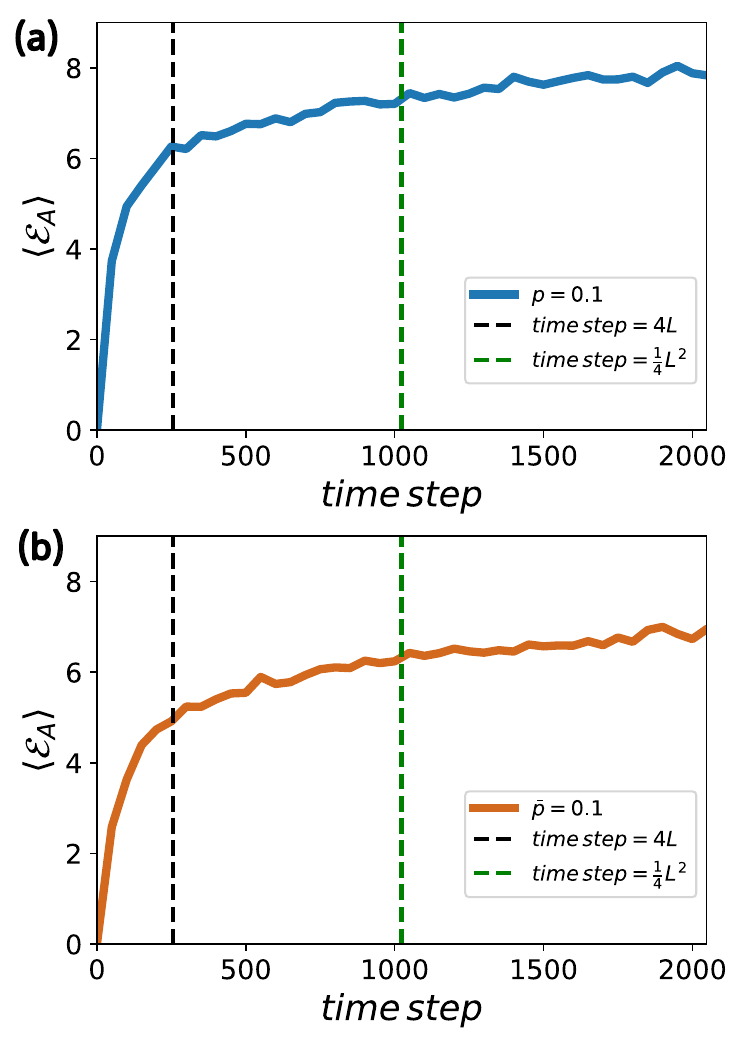}  
\end{center} 
\caption{Typical time evolution behaviors of MBN for uniform probability case I [(a)] and spatially modulated probability case II [(b)].}
\label{FigA2}
\end{figure}
Consequently, from the form of the local Hamiltonian $h^{j,j+1}_{NN}$, and spatially modulated local transverse field $h^m_j$, the averaged operation within the small time interval is approximately given by
\begin{eqnarray}
&&(PMP)\cdot(PU_{NN_o}P)\cdot(PMP)\cdot (PU_{NN_e}P)\approx e^{-\delta t H_{\rm eff}}.\nonumber\\
&&H_{\rm eff}\equiv \sum_{j}\biggl(-\frac{2}{5}Z_jZ_{j+1}-\frac{1}{10}Y_jY_{j+1}\biggr)+\sum_{j}\biggr(-\frac{2}{5}-2\gamma_j\biggl)X_j.\nonumber\\
\label{eHam}
\end{eqnarray}
The model is a modified ZY model with a random transverse field. The effective Hamiltonian $H_{\rm eff}$ can give many insights into the late-time state for the circuit considered in the main text.

By accumulating the evolution operator of the small-time interval, the imaginary time dynamics induced by $H_{\rm eff}$ is given by
\begin{eqnarray}
|\rho(t)\rangle\rangle=e^{-t H_{\rm eff}}|\rho(0)\rangle\rangle.
\end{eqnarray}
We set the initial supervector $|\rho(0)\rangle\rangle=\prod_j|\mathcal{I}_j\rangle\rangle$ corresponding to the infinite temperature state $\frac{1}{2^L}I$. 

In many cases, the imaginary time dynamics leads to a ground state of the Hamiltonian of $H_{\rm eff}$, 
$$
\lim_{t\to \infty}|\rho(t)\rangle\rangle = |GS\rangle\rangle.
$$
where $|GS\rangle$ is a groundstate of the Hamiltonian $H_{\rm eff}$. This ground state gives qualitative insight to the late-time state in the Clifford circuit considered in the main text.\\

\section*{Appendix B: Time evolution and saturation behavior of MBN} 
In this Appendix, we show typical time-evolution and observe saturation behaviors of MBN for both uniform and spatially modulated probability cases, where we set $p=0.1$ and $\bar{p}=0.1$.
The numerical results are displayed in Fig.~\ref{FigA2}. These results indicate that taking the time step $\mathcal{O}(L^2)$ is sufficient to get near steady states while for $\mathcal{O}(L)$ times step may not. 
We expect that this tendency is true for different $p$'s or $\bar{p}$'s except for just critical transition points. In addition, by taking account into the limit of our numerical resource, we observe a late-time state at $t_{step}=L^2/4$ in most of numerical calculations in this work.

\endwidetext
\bibliography{ref}
\end{document}